\documentclass[11pt]{amsart}
\usepackage{geometry}                
\usepackage{graphicx}
\usepackage{amssymb}
\usepackage{epstopdf}
\title{Locating earthquake epicenter without a seismic velocity model}
\author{Rong Qiang Wei}
\address{College of Earth and Planetary Sciences, University of Chinese Academy of Sciences, Beijing, PRC, 100049}
\email{wrq1973@ucas.edu.cn}
\date{}

\begin{document}
\maketitle

\begin{abstract}
 We present a method for locating the seismic event epicenters without assuming an Earth model of the seismic velocity structure, based on the linear relationship between $\log R$ and $\log t$ (where $R$ is the radius of spherical P wave propagated outwards from the hypocenter, $t$ is the travle-time of the P wave). This relationship is derived from the dimensional analysis and a lot of theoretical or real seismic data, in which the earthquake can be considered to be a point source. Application to 1209 events occurred from 2014 to 2017 in the IASPEI Ground Truth (GT) reference events list shows that our method can locate the correct seismic event epicenters in a simple way. $\sim 97.2$\% of seismic epicenters are located with both longitude and latitude errors $\in[-0.1^\circ, +0.1^\circ]$. This ratio can increase if with a finer search grid. As a direct and global‐search location, this method may be useful in obtaining the earthquake epicenters occurred in the areas where the seismic velocity structure is poorly known, the starting points or the  constraints for other location methods.     
\end{abstract}

{\hspace{2.2em}\small Keywords:}

{\hspace{2.2em}\tiny earthquake epicenter, epicentral distance, P waves, travle-time, dimensional analysis}

\section{Introduction}

Earthquake epicenter is one of the basic parameters and the first step in a wide field of seismology.  A lot of methods are developed to locate the epicenters of the seismic events, and a good review can be found in Lomax et al. (2014). Most of these methods, either the deterministic or the probabilistic, rely on a seismic velocity model of the Earth. This results in that the velocity model selected has a strong influence on the locating epicenters of seismic sources (e.g., Kennett, 1992).  Either one-dimensional (1D) earth models, such as PREM (Dziewonski and Anderson, 1981), IASPEI91 (Kennett and Engdahl, 1991), and ak135 (Kennett et al., 1995), or the three-dimensional (3D) velocity models, are used in seismic location (eg., Lomax et al., 2000; Bond$a$r et al., 2018). With these models, a large number of earthquake epicenters have been located. Some famous Bulletins are compiled, such as the PDF Bulletin (The Preliminary Determination of Epicenters), the ISC Bulletin (International Seismological Center), the ISC-EHB Bulletin, and so on. These epicenters play an important role in determining plate (and/or block) boundaries, inferring the distribution of the insidious faults, disaster prevention and relief. 

However,  firstly we may not know the seismic velocity structure well. One of the most obvious example is the existence of strong coda in most recordings of earthquakes, which indicates the strong scattering from unknown structures (Zhao and Curtis, 2019). Secondly many problems are related to the Earth's seismic velocity structure. For example, Wuestefeld et al. (2018) asked whether if the Earth's structure reliably could be approximated by a 1D velocity model or even a 3D model above, and how many layers are available. Thirdly in some areas the seismic velocity structure is poorly known at all. The inaccurate velocity structure can result in incorrect location of the epicenters, and vice versa, because there are tradeoffs between the velocity structure and the location. 

Regardless of the seismic velocity structure, it is well known that many event location methods are based on the linearization approach of Geiger (1910). These methods minimize the residuals between observed and theoretical phase arrival times relating to possible source locations through Taylor expansion involving the first order partial derivatives.  These schemes can accurately locate the Earth's epicenter if and only if with good initial values.  For example, the improved location procedures at the ISC employ a non-linear grid search method to obtain initial hypocentre guess for its linearized location algorithm (Bond$\acute{a}$ and Storchak, 2011). For the other non-linear event location methods, good starting points from other methods are necessary. At the same time,  to estimate the theoretical arrival times of a phase, we have to go back to the seismic velocity model again.

Hence it is necessary to develop a method to obtain the Earthquake epicenter independent of the seismic velocity model.  Andson (1981) presented the arrival order or bisector method, from which the epicenters can be obtained without a seismic velocity model.  This method was developed further  (eg., Nicholson et al. 2004; Rosenberger 2009), and it is useful for starting locations for linearized methods above, or obtaining some constraints on the location of events far outside of an observing station network. 

Here we present another method to provide constraints or the initial values independent of an Earth's seismic velocity structure. In the following sections, we will : (i) analyze the variation of $R$ with $t$ (where $R$ is the radius of spherical P wave propagated outwards from the hypocenter, $t$ is the travle-time of the P wave) with the dimensional analysis; (ii) infer the actual $R\sim  t$ relationship from some theoretical or real seismic data; and (iii) use this $R\sim t$ to locate the seismic epicenters. Finally, some related problems will be discussed further.

\section{Dimensional analysis on the earthquake and a linear relationship between $\log R\sim \log t$}

If the mathematical model or equation of a physical system is unknown, the dimensional analysis can be used to analyze it and its control parameters, design experiments, and then find out the quantitative relationship between the main physical quantities. A famous example is worked out for estimating the energy of a nuclear explosion by Taylor (1950). One of his results states that a spherical shock wave is propagated outwards whose radius $R$ is related to the time $t$ since the explosion started with $R\sim t^{2/5}$.  

The principle of the dimensional analysis is the dimensional homogeneity of the physical quantities. That is to say that the left-hand side of any physical equation must have dimensions like those of the right-hand side. The key concept in dimensional analysis is the systematic identification of dimensionless variables known as $\Pi$ groups. The fundamental theorem in dimensional analysis is the Buckingham’s $\Pi$ theorem, which states that any physical relationship between $n$ variables involving $k$ fundamental physical dimensions (eg., mass, length, and time) can be reduced to a relationship among $n-k$ dimensionless variables. 

Similar to Taylor (1950), we apply this tool of dimensional analysis to an earthquake for the P spherical wave to study its grow radius, $R$.  We assume that the earthquake occurs in the elastic Earth. The relevant physical variables in this problem involves: the characteristic scale $l$ of the earthquake, the energy $E$ released from the earthquake, the travel time $t$ of P wave, the density $\rho$ of the earth media, Young's modulus $E_{_y}$ of elasticity, and Poisson's ratio $\lambda$.  $R$ is a function of $l$, $E$, $t$, $\rho$, $E_{_y}$, and $\lambda$, 

\begin{equation}\label{eq1}
R=f(l,E,t,\rho,E_{_y},\lambda)
\end{equation}
where $f(\cdot)$ represents an unknown function.

Eq. (\ref{eq1}) has a dimensionless variable $\lambda$ and six variables involving various combinations of three mechanical fundamental physical dimensions: mass, length, and time.  According to Buckingham's $\Pi$ theorem,  these variables must be related in a manner that can be expressed by $7-3 =4$ independent dimensionless variables. Selecting $t$, $E$, $\rho$ as the fundamental physical variables, Eq. (\ref{eq1}) can be recast in the dimensionless form,

\begin{equation}\label{eq2}
\frac{R}{t^{2/5}\rho^{-1/5}E^{1/5}}=f(\frac{l}{t^{2/5}\rho^{-1/5}E^{1/5}},\frac{E_{_y}}{t^{-6/5}\rho^{3/5}E^{2/5}},\lambda)
\end{equation} 

Since $$\frac{l}{t^{2/5}\rho^{-1/5}E^{1/5}}/\frac{R}{t^{2/5}\rho^{-1/5}E^{1/5}}=\frac{l}{R}$$ 


Eq. (\ref{eq2}) can be transformed as,


\begin{equation}\label{eq3}
\frac{R}{t^{2/5}\rho^{-1/5}E^{1/5}}=f(\frac{l}{R},\frac{E_{_y}}{t^{-6/5}\rho^{3/5}E^{2/5}},\lambda)
\end{equation}

Generally $l<<R$, therefore $l/R$ has no influence on the $R/t^{2/5}\rho^{-1/5}E^{1/5}$, Eq. (\ref{eq3}) can be transformed further as,

%

\begin{equation}\label{eq4}
\frac{R}{t^{2/5}\rho^{-1/5}E^{1/5}}=f(\frac{E_{_y}}{t^{-6/5}\rho^{3/5}E^{2/5}},\lambda)
\end{equation}

Dimensional analysis has to stop because no more information can be obtained further, and the form of $f(\cdot)$ can only to be determined by the experiment or other approaches. Based on a lot of "experiments", and considering that $E_{_y}$, $\rho$ and $\lambda$ are not constants for the real Earth,  we infer $f(\cdot)= f_1({E_{_y}}/{t^{-6/5}\rho^{3/5}E^{2/5}})f_2(\lambda)$, and $f_1(x)\approx x^\alpha$ (where $\alpha$ is any real number). That is to say, there is a linear relationship between $\log R\sim \log t$, and,

\begin{equation}\label{eq5}
\log R=(2/5+6\alpha/5)\log t+\log\left[\rho^{-1/5-3\alpha/5}E^{1/5-2\alpha/5}{E_{_y}}^\alpha f_2(\lambda)\right]
\end{equation}
where $(2/5+6\alpha/5)$ and $\log\left[\rho^{-1/5-3\alpha/5}E^{1/5-2\alpha/5}{E_{_y}}^\alpha f_2(\lambda)\right]$ can be constants for a particular earthquake but vary with different earthquakes. We can found this from the following "experiments".

Our "experiment" data of $R$ and $t$ is from theoretical calculations and actual earthquake databases, respectively. Due to the space limitations of this paper, we only show randomly the results of 15 earthquakes.  Here and later, $R$ is calculated in the ECEF (earth-centered, earth-fixed) cartesian coordinate system. It should be noted that here $R$ is not the epicentral distance or hypocentral distance, although the calculation method may be the same for local earthquakes.  

\begin{figure}[htb]
 \includegraphics[scale=0.5]{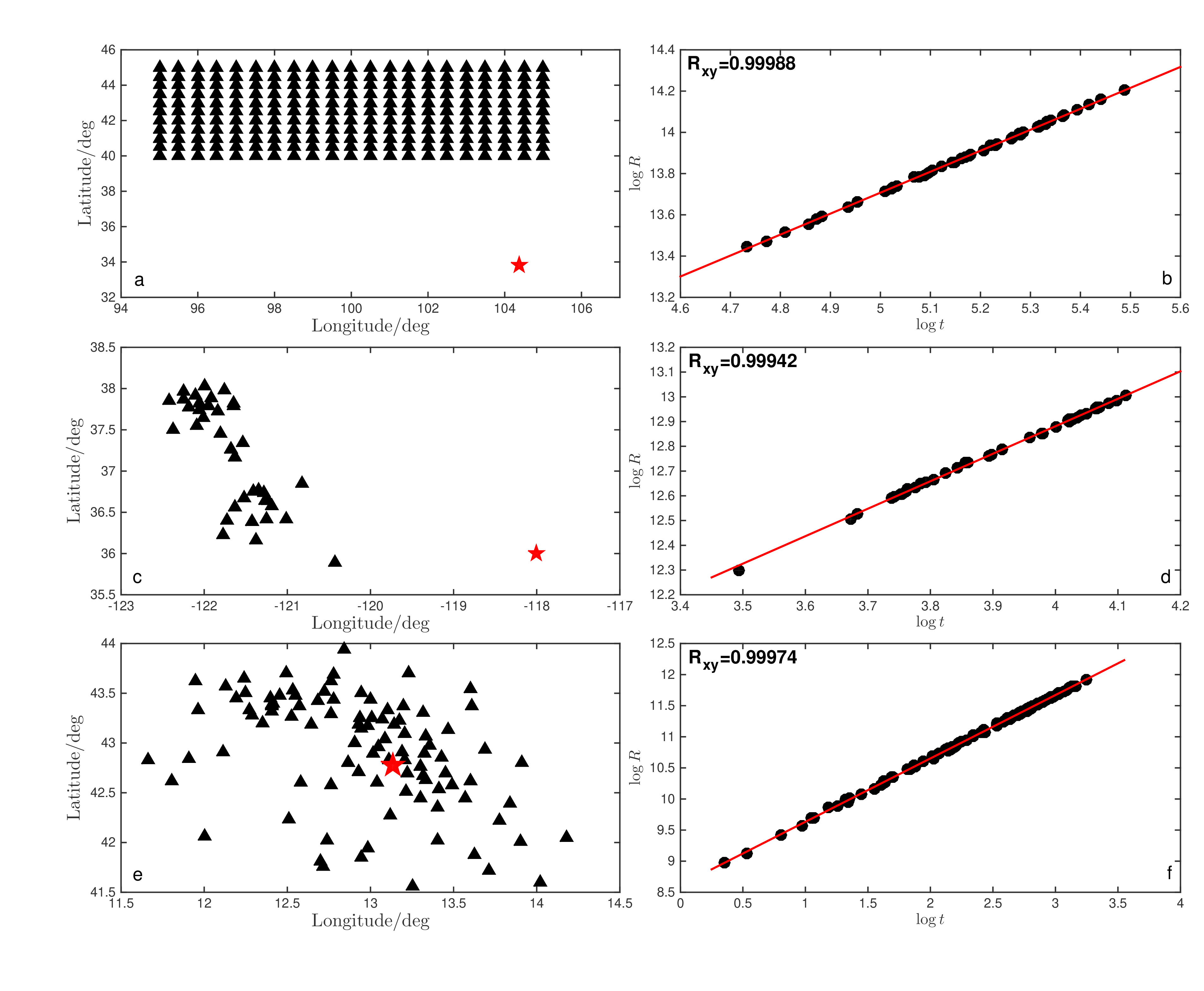}
 \caption{\footnotesize  Subfigure a, c, and e show the epicenter (star) and stations (triangles) for three earthquakes, respectively; Subfigure b, d, and f are the corresponding  $\log R \sim \log t$ for these earthquakes, respectively. $R$ is radius of spherical P wave propagated outwards from the hypocenter and calculated in the ECEF (earth-centered, earth-fixed) cartesian coordinate system, $t$ is for Pg wave and is calculated with Regional Seismic Travel Time (RSTT) software package (Myers et al., 2010). The depths for the earthquake are at 5 km.  The solid lines in the subfigure b, d, and f are the fitting lines. $R_{xy}$ is the linear correlation coefficient.}
\label{fig1}
\end{figure}

In Fig. \ref{fig1}, the travel-times $t$ of three earthquakes are calculated with Regional Seismic Travel Time (RSTT) software package (Myers et al., 2010);  Data of epicenters and stations are selected randomly.  We calculated $t$ for Pg wave in three cases: (1) The stations distribute regularly; (2) The stations distribute irregularly on one side of the epicenter; (3) The stations distribute irregularly around the epicenter. The depths for the earthquake are at 5 km.  It can be found that $\log R $ of these three earthquakes increases linearly with increasing $\log t$; The linear correlation coefficients are greater than 0.99. Case (1) has the relative maximum of correlation coefficient, and case (2) has the relative minimum of correlation coefficient.

\begin{figure}[htb]
 \includegraphics[scale=0.5]{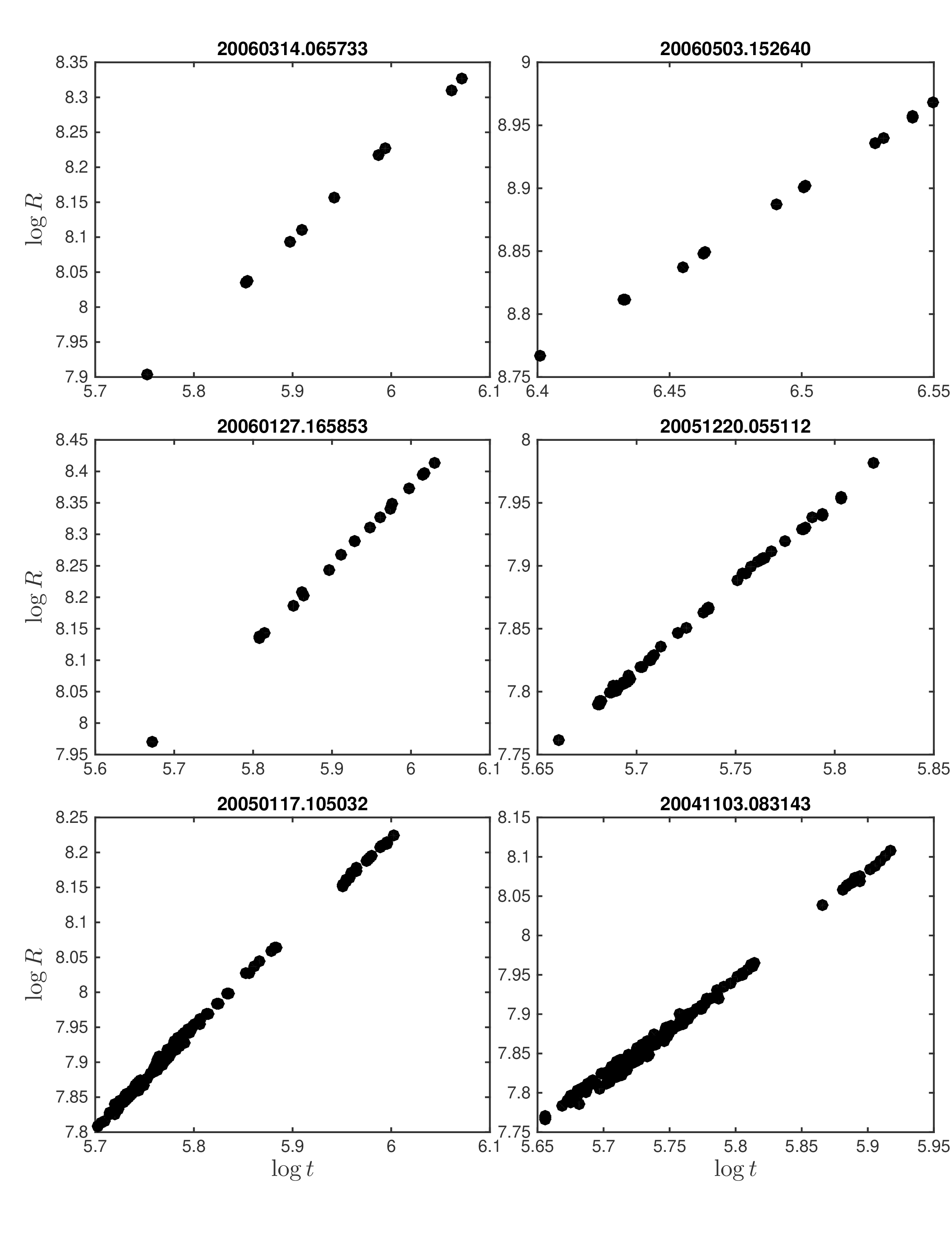}
 \caption{\footnotesize  $\log R \sim \log t$ for the earthquakes from an Japanese database of seismic traveltimes (Yoshimitsu and Obayashi, 2017). The title of each subfigure is the code for the earthquake. $R$ is calculated in the ECEF (earth-centered, earth-fixed) cartesian coordinate system. }
\label{fig2}
\end{figure}

\begin{figure}[htb]
 \includegraphics[scale=0.5]{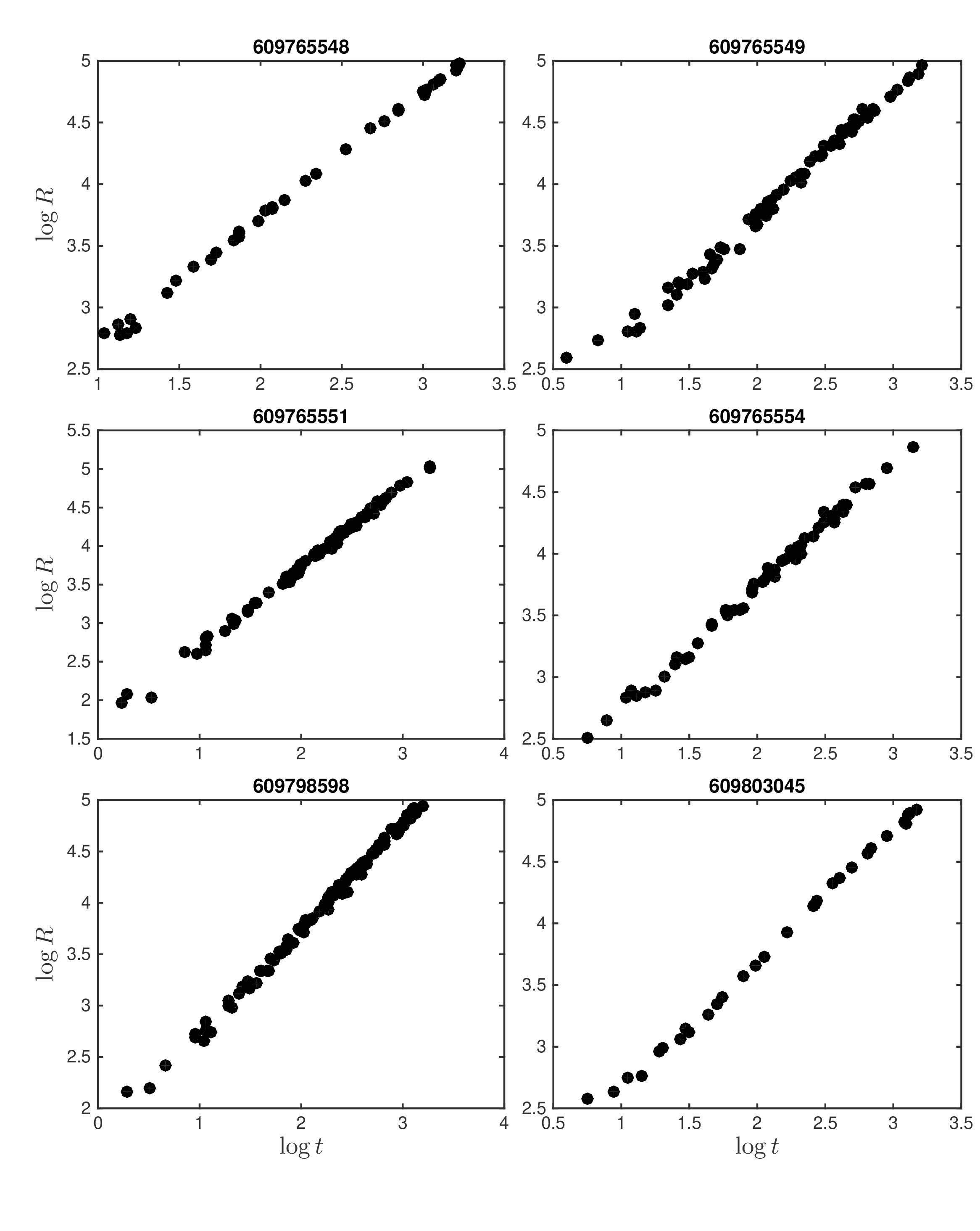}
 \caption{\footnotesize  $\log R \sim \log t$ for the earthquakes from the IASPEI Ground Truth (GT) reference events list (Bond$\acute{a}$r et al., 2004, 2008; Bond$\acute{a}$r and McLaughlin, 2009). $R$ is calculated in the ECEF (earth-centered, earth-fixed) cartesian coordinate system. The title of each subfigure is the unique id number specific to each IASPEI reference event .}
\label{fig3}
\end{figure}

In Fig. \ref{fig2} and Fig. \ref{fig3}, data of $t$ for P wave,  hypocenter parameters (epicenters and depth) and stations are from actual earthquakes. Those data in Fig. \ref{fig2} are from an Japanese database of seismic traveltimes, in which the absolute arrival times of seismic phase P was measured by manual picking and differential travel times between pP and P were obtained using waveform cross-correlation method (Yoshimitsu and Obayashi, 2017).  Those data in Fig. \ref{fig3} are from the IASPEI Ground Truth (GT) reference events list\footnote{International Seismological Centre (2019), IASPEI Reference Event (GT) List, https://doi.org/10.31905/32NSJF7V} (Bond$\acute{a}$r et al., 2004, 2008; Bond$\acute{a}$r and McLaughlin, 2009). For the sake of simplicity, the fitting lines are not shown in Fig. \ref{fig2} and Fig. \ref{fig3}. It can also be found that $\log R $ of these earthquakes increases linearly with increasing $\log t$. 

\begin{figure}[htb]
 \includegraphics[scale=0.4]{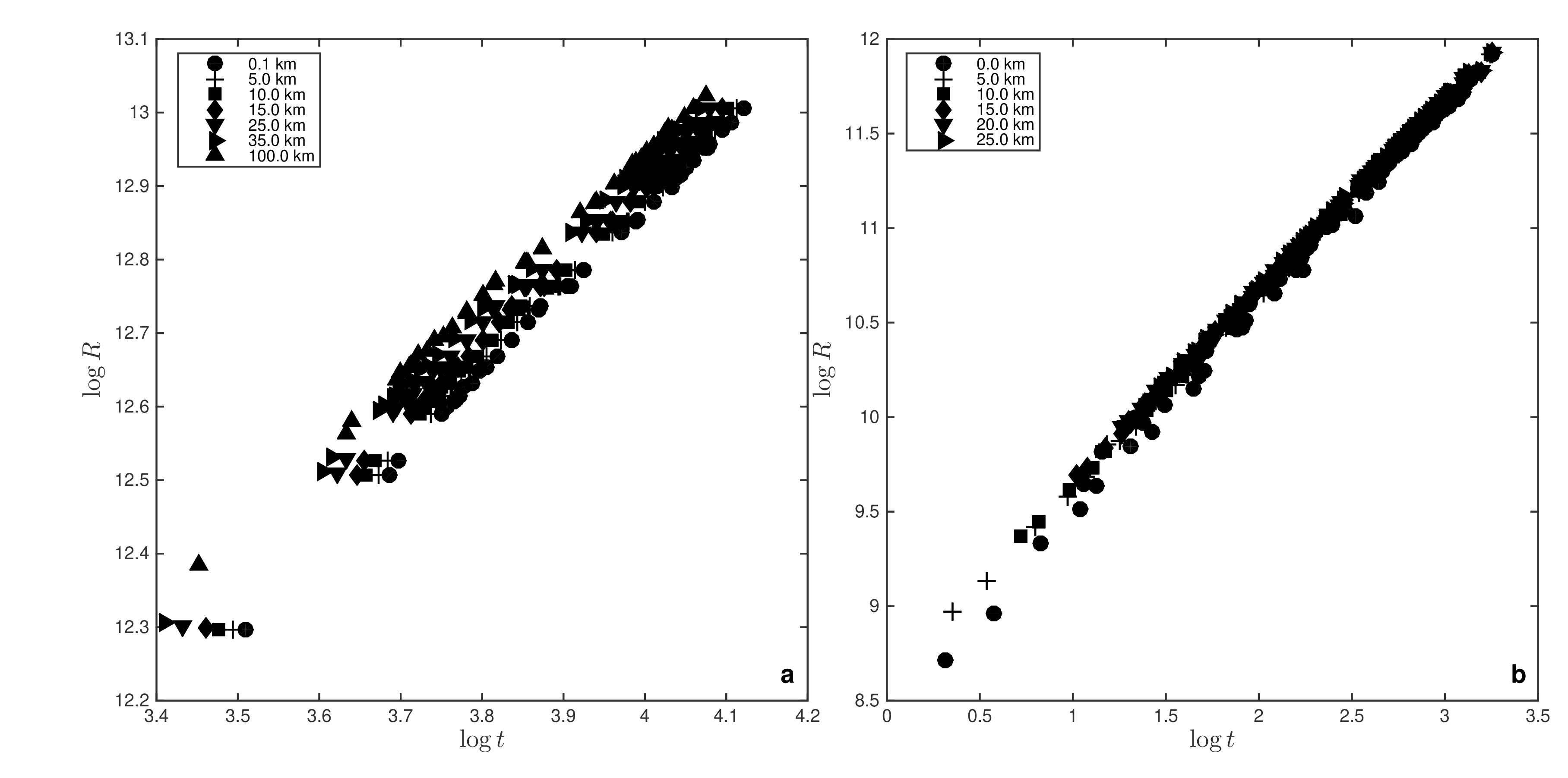}
 \caption{\footnotesize  (a) $\log R \sim \log t$ for the earthquake in Fig. \ref{fig1} c but at different depths. (b) $\log R \sim \log t$ for the earthquake in Fig. \ref{fig1} e but at different depths. $R$ is calculated in the ECEF (earth-centered, earth-fixed) cartesian coordinate system. $t$ is for Pg wave and is calculated with RSTT software package.}
\label{fig4}
\end{figure}

In Fig. \ref{fig4},  $t$ is again calculated for the earthquakes at different depths with RSTT software package (Myers et al., 2010);  Data of epicenters and stations are the same to the case (2) and (3) in Fig. \ref{fig1}.  It can be seen that: (1) At a fixed depth, $\log R $ of these earthquakes increases again linearly with increasing $\log t$ (2) With the increasing depth, these lines move nearly parallel to the left; This movement in Fig. \ref{fig4}a is more obviously than that in Fig. \ref{fig4}b.  This means that the variation in depth only affects the intercepts of the lines, but does not affect the slope of the lines.   However, such effects are not significant, especially those in the Fig. \ref{fig4}b. We will mention this in the later discussions.

Furthermore, it can also be found that the slopes of $\log R \sim \log t$ lines are different. In Fig. \ref{fig1}, the slopes are 1.017, 1.110, 1.019 for case (1), case (2), and case (3), respectively; In Fig. \ref{fig2}, the slopes have a mean of 1.328; In Fig. \ref{fig3}, the slopes have a mean of 1.025;  In Fig. \ref{fig4}, the slopes decrease with increasing depth from 1.124 to 1.005 for the earthquake in subfigure (a), and from 1.061 to 0.994 for the earthquake in subfigure (b). Therefore, the slopes of $\log R \sim \log t$ lines vary with different earthquakes.

In summary, these "experiments" show that there is a linear relationship between $\log R\sim \log t$. Since the intercepts and the slopes of the $\log R\sim \log t$ lines are different from one earthquake to another earthquake,  we infer that $(2/5+6\alpha/5)$ and $\log\left[\rho^{-1/5-3\alpha/5}E^{1/5-2\alpha/5}{E_{_y}}^\alpha f_2(\lambda)\right]$ can be constants for a particular earthquake but vary with different earthquakes.

\section{A method to locate earthquake epicenter without a seismic velocity model}

Eq. (\ref{eq5}) can be used to locate the earthquake. In the ECEF cartesian coordinate system, we can translate this equation into Eq. (\ref{eq6}) in the following,

\begin{equation}\label{eq6}
\frac{1}{2}\log\left[(x-x_0)^2+(y-y_0)^2+(z-z_0)^2\right] =a\log (t_s-t_0)+b
\end{equation} 
where $a=(2/5+6\alpha/5)$ and $b=\log\left[\rho^{-1/5-3\alpha/5}E^{1/5-2\alpha/5}{E_{_y}}^\alpha f_2(\lambda)\right]$,$x_0,y_0,z_0,t_0$ are hypocenter parameters. $x,y,z,t_s$ are the location coordinates of the stations and arrive time, respectively. $a, b$ are two constants for a particular earthquake. Then the travel time of the P wave $t=t_s-t_0$. 

There are 6 unknown variables in Eq. (\ref{eq6}), and they can be solved with the location coordinates of 6 stations in theory. However, it can be seen that the system of equations is highly nonlinear, it can not be solved easily. In fact, we tried many methods and no any available and stable solution was obtained.  

A lot of tests show that we can just use the linear relationship between $\log R$ and $\log t$ only to locate $x_0,y_0$, while $z_0$ can not be located by this way which can be seen from Fig. \ref{fig4}, and we will discuss this in the following section. In our scheme, the key is to find out the maximum of the linear correlation coefficient between $\log\left[(x-x_0)^2+(y-y_0)^2+(z-z0)^2\right]$ and $\log t$ (Here $z0$ can be fixed at any depth and we let $z_0=5.0$ km. $z$ is the elevation of the station). The corresponding longitude and latitude to $(x_0, y_0)$ are the epicenter. Because we can adopt simple grid search, the epicenter obtained here is a direct and global‐search location.     

\begin{figure}[htb]
 \includegraphics[scale=0.32]{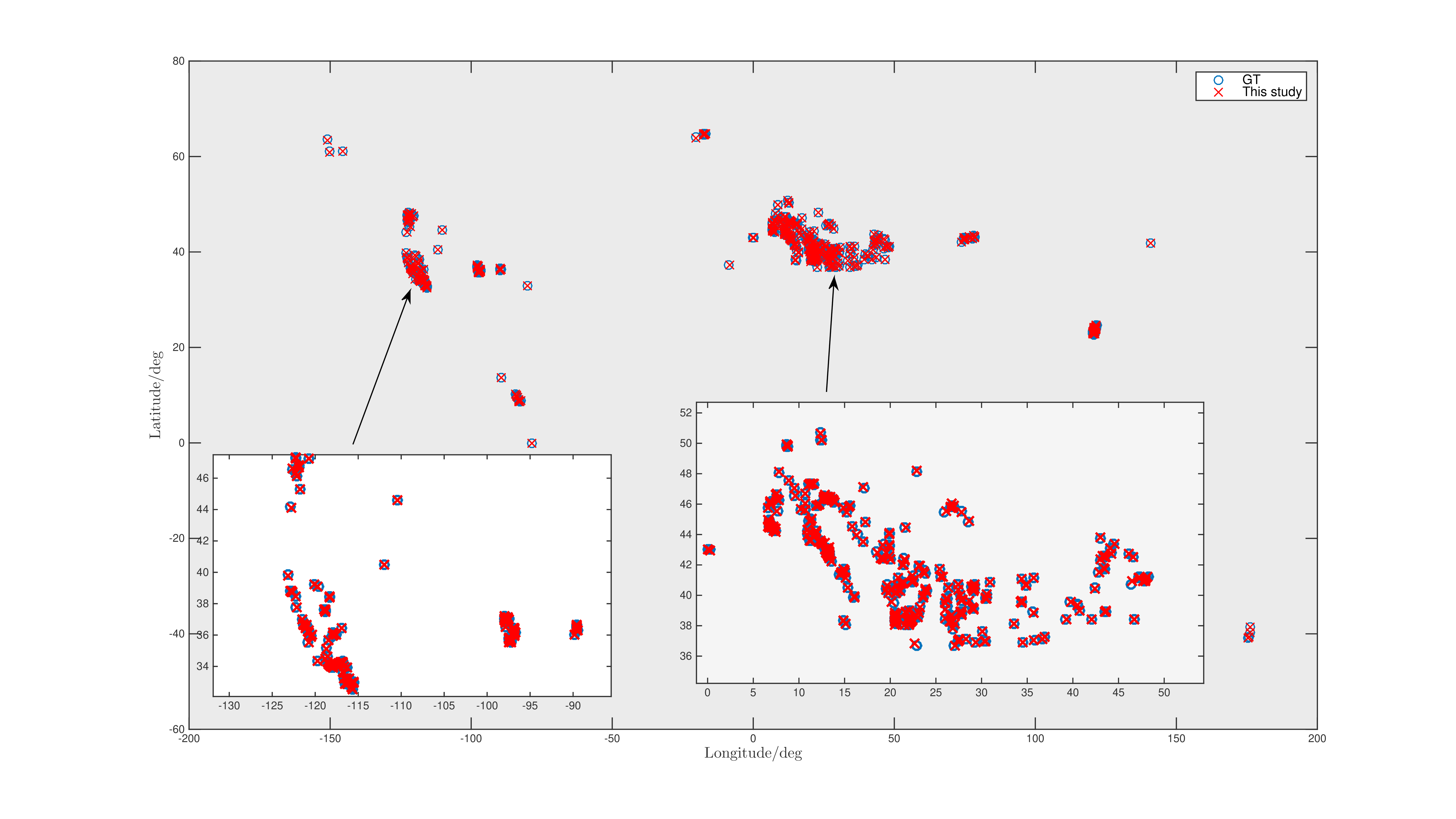}
 \caption{\footnotesize Comparison of the epicenters for 1209 seismic events occurred from 2014 to 2017 from our location method with those in the IASPEI GT reference events list. }
\label{fig5}
\end{figure}

\begin{figure}[htb]
 \includegraphics[scale=0.33]{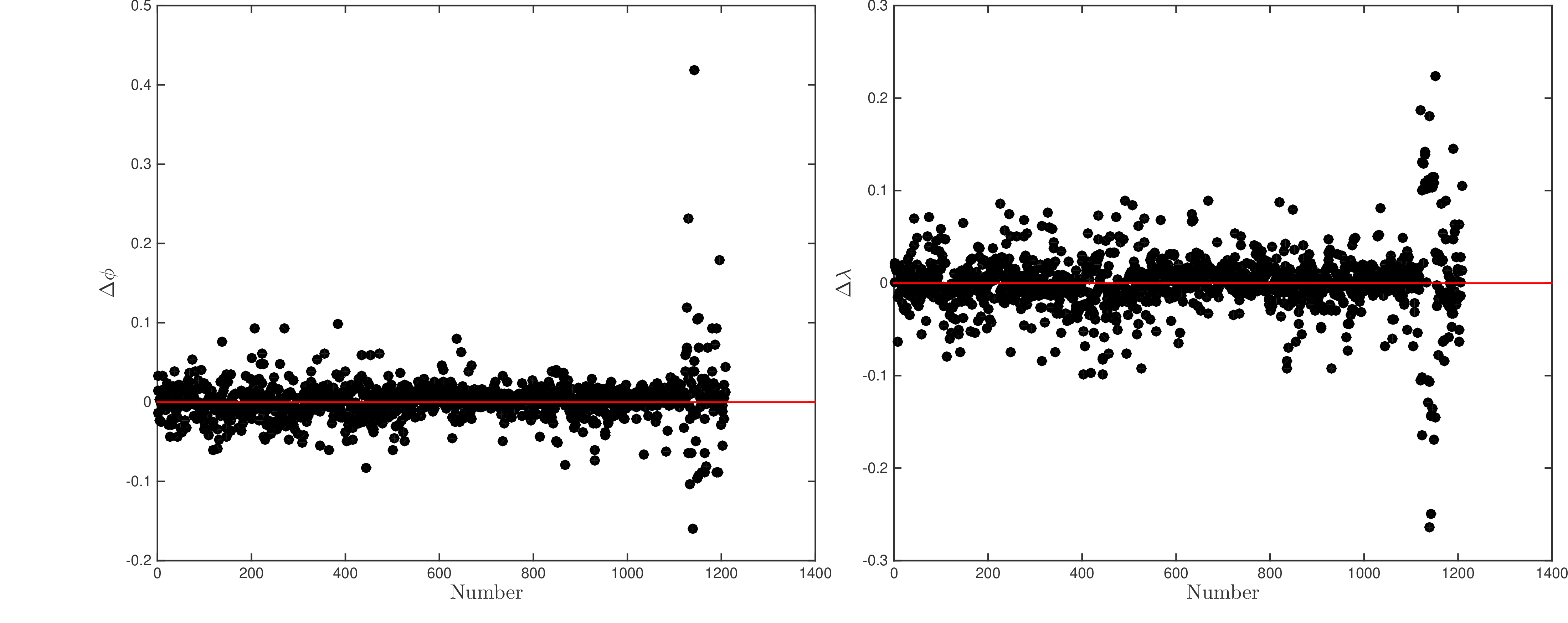}
 \caption{\footnotesize Errors in the latitude (a) and longitude (b) between the epicenters from our method and those in the GT reference events list. }
\label{fig6}
\end{figure}

We apply this method to 1209 seismic events occurred from 2014 to 2017 in the IASPEI Ground Truth (GT) reference events list.  A grid size of $0.01^\circ$ is adopted here.  Fig. \ref{fig5} shows our location results.  It can be seen that the epicenters from our method are consistent well with those in the IASPEI GT reference events list. Fig. \ref{fig6} shows the corresponding errors of the longitudes and latitudes, respectively. It can be found that most absolute values of these errors are less than $0.1^\circ$. Further statistics show that there are 1175 seismic events with both longitude and latitude errors $\in[-0.1^\circ, +0.1^\circ]$, while those of 1056 events $\in[-0.05^\circ, +0.05^\circ]$. That is to say,  about 97.2\% of seismic epicenters are located with both longitude and latitude errors $\in[-0.1^\circ, +0.1^\circ]$, while about 87.3\% epicenters are located with errors $\in[-0.05^\circ, +0.05^\circ]$. An excepted latitude error is $ 0.42^\circ$. However, it should be pointed out that (1) If a smaller grid size, e.g., $0.001^\circ$ or $0.0001^\circ$, is used, the errors above will be reduced but it will spend more compution time; (2) The linear correlation coefficients calculated from our epicenters are always greater than those calculated with epicenters from the GT reference events list. For example, for the earthquake with above latitude error of $ 0.42^\circ$, the linear correlation coefficient is 0.9387 from the GT epicenter, but it is 0.9985 from that located by us. 

\section{Discussions and Conclusions}

Finding new methods for locating or constraining the earthquake epicenters remains an active area of research. Especially for the cases of the explosion, landslide, rockslide, snow avalanche, or ice quake in a wide field of engineering, we only need the epicenters. The tests in the previous section show that the epicenter ($x_0,y_0$) can be located just only using the linear relationship between $\log R$ and $\log t$ without any seismic velocity structure. Because this linear relationship is not proved in theory but inferred from "experiments", this scheme is an empirical one. Despite this, as it can be seen from previous section it can locate the correct earthquake epicenters. This method is easy to implement for the key is only to find out the maximum of the linear correlation coefficients between $\log R$ and $\log t$.  Hence this new method can be used as an independent method to locate the epicenter, or it can provide another initial values or constraints for other methods depend on the Earth's seismic velocity model.

\begin{figure}[htb]
 \includegraphics[scale=0.6]{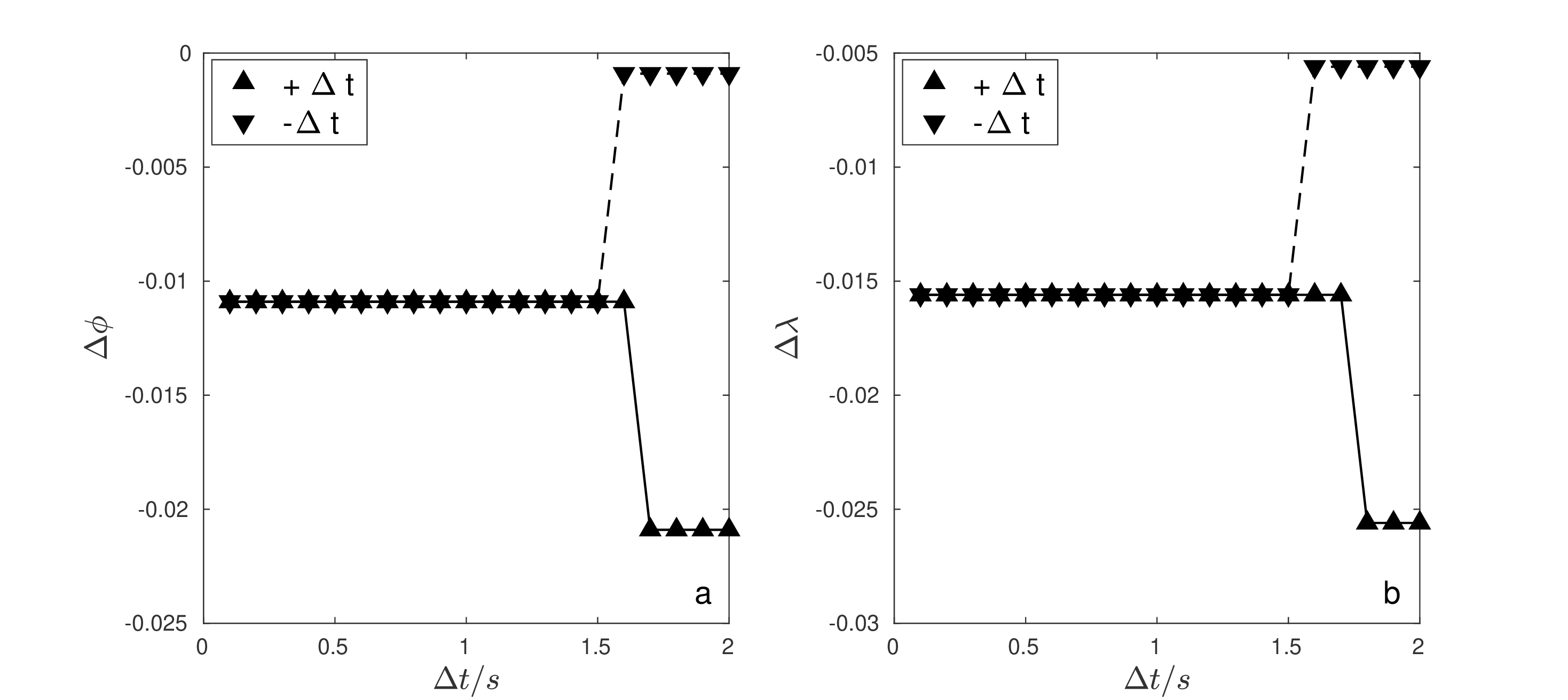}
 \caption{\footnotesize Errors for the epicenter latitude (a) and longitude (b) result from the errors of $t$. $+\Delta t $ represent a delay in travel time, while $-\Delta t $ represent an advance in travel time. The earthquake is that in Fig. \ref{fig1} e at a hypocentral depth of 15.0 km. $R$ is calculated in the ECEF (earth-centered, earth-fixed) cartesian coordinate system. $t$ is for Pg wave and is calculated with RSTT software package. }
\label{fig7}
\end{figure}

\begin{figure}[htb]
 \includegraphics[scale=0.4]{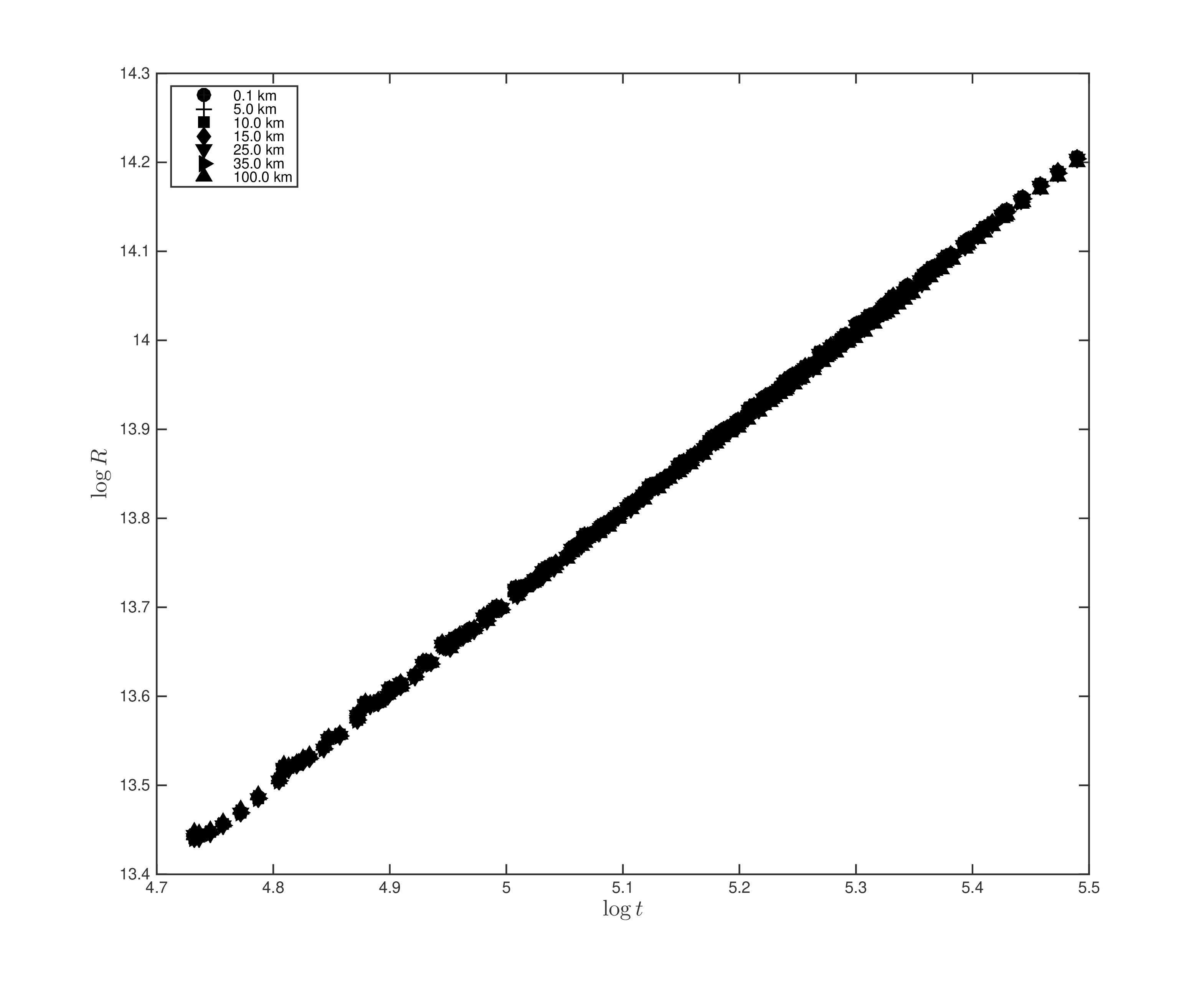}
 \caption{\footnotesize $\log R \sim \log t$ for the earthquakes in Fig. \ref{fig1} a at different hypocentral depths. $R$ is calculated in the ECEF (earth-centered, earth-fixed) cartesian coordinate system. $t$ is for Pg wave and is calculated with RSTT software package. }
\label{fig8}
\end{figure}

\begin{figure}[htb]
 \includegraphics[scale=0.6]{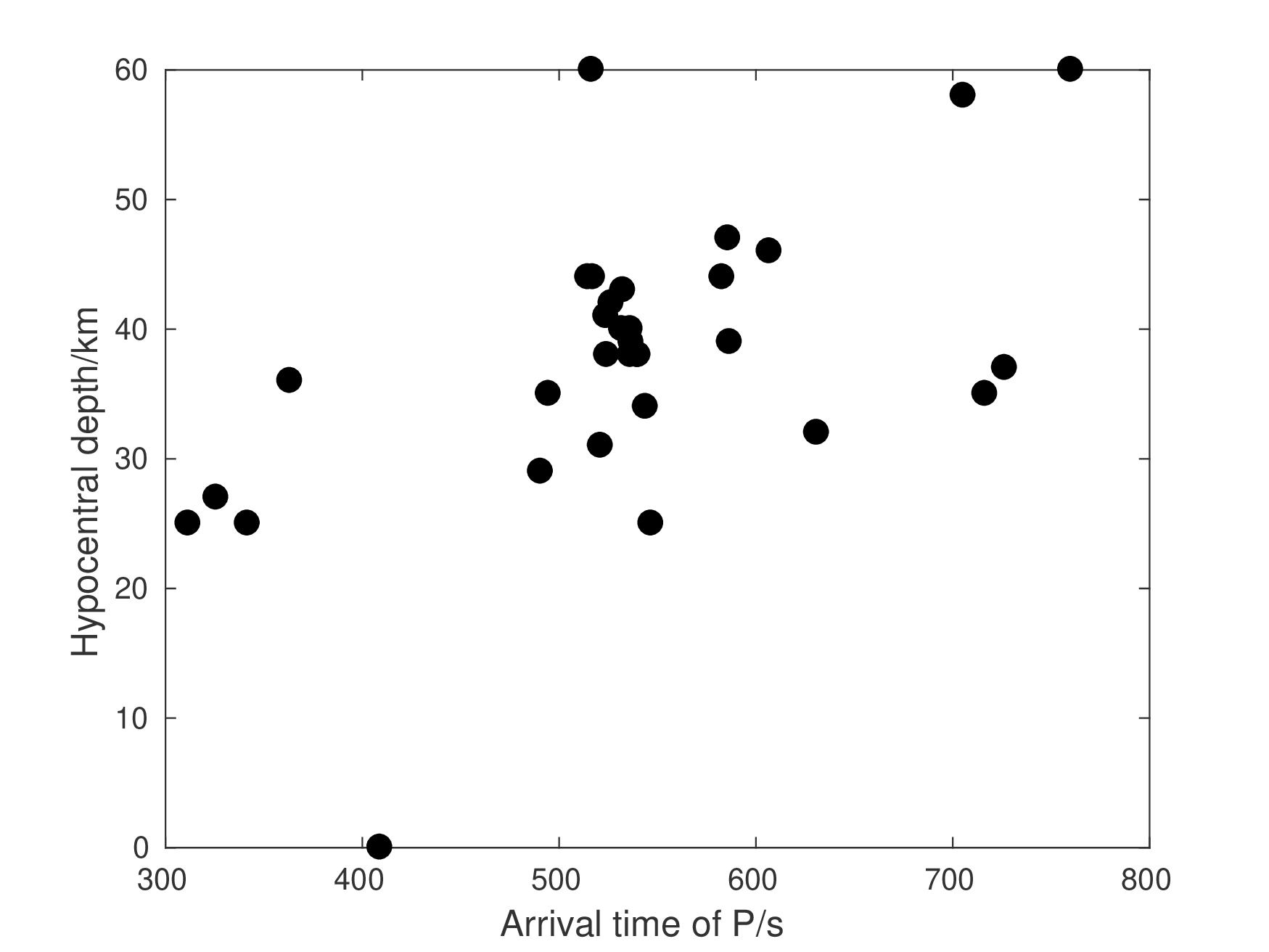}
 \caption{\footnotesize Hypocentral depths vs. travel times for the Event 3160315 in the GT reference events list. Hypocentral depths are estimated from Eq. (\ref{eq7}) with TauP Toolkit (Crotwell et al., 1999). In the estimation, the velocity model for the Earth is the IASPEI91; All the travel times and epicenters are from the GT reference events list. The GT list shows the Event 3160315 occurred at 45 km.}
\label{fig9}
\end{figure}

A possible cause of the error to the epicenter is from the $t$. To check this, we add an error $\Delta t$ from 0.1 s to 2.0 s to the $t$. Fig. \ref{fig7} shows the location errors of the epicenter for the earthquake in Fig. \ref{fig1}e with a hypocentral depth $z_0=15.0$ km. When $\Delta t \le 1.5$ s and whether there is a delay ($+\Delta t $) or an advance ($-\Delta t $) in travel time, it can be seen that location errors are $-0.0109^\circ$ and $-0.0156^\circ$ for latitude and longitude, respectively; And they do not change as the $\Delta t $ increases. When there is an advance in travel time ($-\Delta t $) and $\Delta t \ge 1.6$ s, the latitude errors decrease stepwise from $-0.0109^\circ$ to $-0.0009^\circ$ and the longitude errors decrease stepwise from $-0.0156^\circ$ to $-0.0056^\circ$, respectively, with the increasing $\Delta t $. When there is a delay in travel time ($+\Delta t $) and $\Delta t \ge 1.7$ s (for latitude) or $\Delta t \ge 1.8$ s (for longitude), the latitude errors increase stepwise from $-0.0109^\circ$ to $-0.0209^\circ$ and the longitude errors increase stepwise from $-0.0156^\circ$ to $-0.0256^\circ$, respectively, with the increasing $\Delta t $. Further calculations ($z_0\le 35.0$ km) show that the error distribution varies with the hypocentral depth $z_0$. When $z_0<10.0$ km and when there is an advance in travel time ($-\Delta t $), the errors will oscillate ($z_0\le 5.0$ km) or increase stepwise quickly ($z_0 > 5.0$ km), but the maximum error is less than $0.0400^\circ$ for latitude and $0.0500^\circ$ for longitude, respectively. When $z_0\le 5.0$ km and when there is a delay in travel time ($+\Delta t $), the error increase stepwise but the maximum absolute error is about $0.0100^\circ$ for latitude and $< 0.0300^\circ$ for longitude, respectively. With the increasing $z_0$ ($z_0>10.0$ km), the errors will be negative for both latitude and longitude, but the maximum absolute error is less than $0.0300^\circ$. Roughly speaking, for the earthquake like that in Fig. \ref{fig1}e in the crust, the maximum absolute error for both latitude and longitude of the epicenter caused by the $\Delta t$ is less than $0.0500^\circ$ when $\Delta t\in [0.1, 2.0]\cup [-2.0, -0.1]$ s.    

Unfortunately and obviously, our method can not determine the hypocentral depth $z_0$ of the seismic events. Fig. \ref{fig4}a show that $\log R$ varies linearly with $\log t$ with similar slopes but with different intercepts. Our method use only the linear relationship between $\log R$ and $\log t$, and we do not know which line of $\log R\sim \log t$ in Fig. \ref{fig4}a is used. In fact, the hypocentral depth $z_0$ has little influence on the linear relationship between $\log R$ and $\log t$ relative to the epicenter. This can be seen from the slopes in the Fig. \ref{fig4}a but more clearly from Fig. \ref{fig4}b and Fig. \ref{fig8}. In Fig. \ref{fig8},  $t$ and $\log R $ are calculated at different hypocentral depths with RSTT software package (Myers et al., 2010);  Data of epicenters and stations are the same to those in Fig. \ref{fig1}a. It can be seen that the lines of $\log R\sim \log t$ are indistinguishable, although $\log R $ increases linearly with increasing $\log t$. The reason for this is $(z-z_0)^2 << (x-x_0)^2+(y-y_0)^2$, and $(z-z_0)^2$ has little effect on $R$ unless the station is very close to the epicenter. In the words of travel time, the variation of the hypocentral depth can result in a little change of the travel time relative to that form the epicenter.  Therefore, the $z_0$ can not be determined reliably with this approach. 

However, we can estimate the $z_0$ from a precise epicenter such as that from our method, combining with a reliable velocity structure of the Earth and a precise travel time. This can be found clearly from the equations for the travel time and epicenter when the Earth is spherical and velocity only changes with depth (eg., Buland and Chapman, 1983),

\begin{equation}\label{eq7}
\left\{
\begin{array}{ll}
T(p)&=\int_{r_s}^{r_e}\frac{[u(r)r]^2}{\sqrt{[u(r)r]^2-p^2}}\frac{{\rm{d}}r}{r}\\
\Delta(p)&=\int_{r_s}^{r_e}\frac{p}{\sqrt{[u(r)r]^2-p^2}}\frac{{\rm{d}}r}{r}
\end{array}
\right.
\end{equation}
where $T(p)$ is the travel time, $\Delta (p)$ the epicenter, $p$ the ray parameter, $u(r)$ the slowness, $r_e$ the radius of the Earth, $r_s$ the radius of the hypocentral depth $z_0$ ($r_s+z_0=r_e$).

In Eq. (\ref{eq7}) there are two unknowns $r_s$ and $p$, and they can be obtained when $T(p)$, $\Delta(p)$ and $u(r)$ are known. 
An example is shown in Fig. \ref{fig9} for the Event 3160315 occurred at 45 km in the GT reference events list. The hypocentral depths in this figure are estimated with the TauP Toolkit (Crotwell et al., 1999), which is based on the algorithm of Buland and Chapman (1983). In the calculation, the 1D velocity model for the Earth is the IASPEI91; All the travel times are for P waves; All the travel times and epicenters are from the GT reference events list. It can be found that hypocentral depths $z_0$ estimated are scattered with a mean of 37.8 km and a standard deviation of 11.3 km. These hypocentral depths show that the $z_0$ can be determined from Eq. (\ref{eq7}) when the epicenter, the velocity structure of the Earth and the travel time are precise; But other methods, e.g.,  using readings from the so-called depth phases (e.g., Engdahl et al., 1998) such as pP, should be used for a reliable estimation of the hypocentral depth. 

Finally in our method, an earthquake is taken as a point source, for the assumption that generally $l<<R$ in Eq. (\ref{eq3}). For the case that the station is very close to the epicenter, e.g., Fig. \ref{fig1}e, this assumption can be inaccurate. The corresponding points of $\log R\sim \log t$ will deviate from the line. Therefore, the method here is more reasonable for teleseismic events, or for the case the station is not close the epicenter ($l<<R$ holds). 

\vspace{5em}


\ \ 


\begin{thebibliography}{99}
\bibitem{}
Anderson, K.P., 1981. Epicentral location using arrival time order, Bull. seism. Soc. Am., 30, 119-130.
\bibitem{}
Buland, R., Chapman, C. H., 1983. The Computation of Seismic Travel Times, Bull. Seism. Soc. Am. 73 (5), 1271–1302.
\bibitem{}
Bond$\acute{a}$r, I., Bergman, E. A., Engdahl, E. R., Kohl, B., Kung, Y., and Mclaughlin, K. L., 2008. A hybrid multiple event location technique to obtain ground truth event locations, Geophys. J. Int., 175, https://doi.org/10.1111/j.1365-246X.2011.05011.x
\bibitem{}
Bond$\acute{a}$r, I., Engdahl, E.R., Yang, X.P., Ghalib, H., Hofstetter, A., Kirichenko, V.V., Wagner, R.A., Gupta, I.N., Ekstr$\ddot{o}$m, G.,Bergman, E.A., Israelsson,H., and McLaughlin K.L., 2004. Collection of a reference event set for regional and teleseismic location calibration, Bull. Seismol. Soc. Am., 94, 1528-1545, http://dx.doi.org/10.1785/012003128.
\bibitem{}
Bond$\acute{a}$r, I., M$\acute{o}$nus, P., Czanik, C., Kiszely, M., Gr$\acute{a}$czer, Z., W$\acute{e}$ber, Z., and Alp Array Working Group, 2018. Relocation of Seismicity in the Pannonian Basin Using a Global 3D Velocity Model. Seismological Research Letters, 89(6), 2284-2293.
\bibitem{}
Bond$\acute{a}$r, I. and McLaughlin, K.L., 2009. A New Ground Truth Data Set For Seismic Studies, Seismol. Res. Lett., 80, 465-472, https://doi.org/10.1785/gssrl.80.3.465
\bibitem{}
Bond$\acute{a}$r, I., and Storchak, D., 2011. Improved location procedures at the International Seismological Centre, Geophys. J. Int. (2011) 186, 1220-1244.
\bibitem{}
Crotwell, H. P.,  Owens, T. J. and Ritsema J. , 1999. The TauP ToolKit: Flexible Seismic Travel-Time and Raypath Utilities, Seismological Research Letters. 70 (2), 154–160.
\bibitem{}
Dziewonski, A.M. and Anderson, D.L., 1981. Preliminary Reference Earth Model, Phys. Earth planet. Inter., 25, 297-356.
\bibitem{}
Engdahl, E. R., Van der Hilst, R. D., and Buland, R. P., 1998. Global teleseismic earthquake relocation with improved travel times and procedures for depth determination. Bull. Seism. Soc. Am., 88, 722-743.
\bibitem{}
Geiger, L., 1910. Herdbestimmung bei Erdbeben aus den Ankunftszeiten. Nachrichten von der Königlichen Gesellschaft der Wissenschaften zu G$\ddot{o}$ttingen, Mathematisch-Physikalische Klasse, 331-349. (In 1912 translated to English by Peebles, F. W. L., and Corey, A. H.: Geiger, L. (1912). Probability method for the determination of earthquake epicenters from the arrival time only. Bulletin St. Louis University, 8, 60-71).
\bibitem{}
Kennett, B. L. N., 1992. Locating oceanic earthquakes-the influence of regional models and location criteria. Geophys. J. Int., 108, 848-854.
\bibitem{}
Kennett, B.L.N. and Engdahl, E.R., 1991. Traveltimes for global earthquake location and phase identification, Geophys. J. Int., 105, 429-465.
\bibitem{}
Kennett, B.L.N. and Engdahl, E.R., and Buland, R., 1995. Constraints on seismic velocities in the Earth from travel times, Geophys. J. Int., 122, 108-124. 
\bibitem{}
Lomax A., Virieux J., Volant P., Berge-Thierry C., 2000. Probabilistic Earthquake Location in 3D and Layered Models. In: Thurber C.H., Rabinowitz N. (eds) Advances in Seismic Event Location. Modern Approaches in Geophysics, vol 18. Springer, Dordrecht. Doi: 10.1007/978-94-015-9536-0\_5
\bibitem{}
Lomaxa, A., Michelinib, A and Curtisc, A, 2014. Earthquake Location, Direct, Global-Search Methods, Encyclopedia of Complexity and Systems Science, doi: 10.1007/978-3-642-27737-5\_150-2
\bibitem{}
Myers, S.C., Begnaud, M.L., Ballard, S., Pasyanos, M.E., Phillips,W.S., Ramirez, A.,  Antolik, M., Hutchenson, K.D., Dwyer, J.J., Rowe,C.A., Wagner, G.S.,, 2010. A crust and upper-mantle model of Eurasia and North Africa for Pn travel-time calculation. Bull. Seismol. Soc. Am., 100(2), 640-656. doi: 10.1785/0120090198.
\bibitem{}
Nicholson, T., Gudmundsson, $\acute{O}$., Sambridge, M., 2004, Constraints on earthquake epicentres independent of seismic velocity models. Geophys J Int, 156: 648–654 
\bibitem{}
Rosenberger, A., 2009. Arrival-time order location revisited. Bull. Seism. Soc. Amer., 99, 2027-2034.
\bibitem{}
Taylor, G., 1950. The Formation of a Blast Wave by a Very Intense Explosion. I. Theoretical Discussion, Proc. R. Soc. Lond. A, 201, 159-174
\bibitem{}
Wuestefeld, A.,Greve, S. M., N$\ddot{a}$sholm, S.P., and Oye, V., 2018. Benchmarking earthquake location algorithms: A synthetic comparison, Geophysics, 83(4), KS35–KS47. doi:10.1190/GEO2017-0317.1
\bibitem{}
Yoshimitsu, J., Obayashi, M., 2017. A database of global seismic travel times, JAMSTEC Rep. Res. Dev., 24, 23-29
\bibitem{}
Zhao, Y., and Curtis, A., 2019. Relative source location using coda-wave interferometry: Method, code package, and application to mining-induced earthquakes. Geophysics, 84(3), F73-F84.
\end{thebibliography}
\end{document}